\input{aipcheck}

\documentclass[
  ,draft            % use draft while you are working on the paper
  ]
  {aipproc}

\layoutstyle{6x9}

\begin{document}

\title{Cosmic Ray Feedback}



%\classification{<Replace this text with PACS numbers; choose from this list:
%                \texttt{http://www.aip..org/pacs/index.html}>}
%\keywords      {<Enter Keywords here>}
%\classification{missing classification}
%\keywords{Cosmic Ray Feedback}

\author{William G. Mathews}{
  address={Department of Astronomy and Astrophysics, UC Santa Cruz}
}

%\author{<author2>}{
%  address={<common address for author2 and author3>}
%}

%\author{<author3>}{
%  address={<common address for author2 and author3>}
%  ,altaddress={<author1 address>} % additional visiting address
%}

\begin{abstract}
% This template file shows how to use the \texttt{aipproc} class to
% produce a paper with the correct layout for \emph{%
%   AIP Conference Proceedings  6in   x 9in single column}.
%
% A full description of the features supported by the \texttt{aipproc}
% class can be found in the \texttt{aipguide.pdf} document accompanying
% the distribution.
%
% Frequently asked questions can be found in the \texttt{FAQ.txt}
% document.
Cosmic rays produced or deposited at sites in hot cluster gas 
are thought to provide the pressure that forms X-ray cavities.
While cavities have a net cooling effect on cluster gas, 
young, expanding cavities drive shocks that increase the local 
entropy.
Cavities also produce radial filaments of thermal gas and 
are sources of cluster cosmic rays that diffuse through cavity walls, 
as in Virgo where a radio lobe surrounds a radial 
thermal filament.
Cosmic rays also make the hot gas locally buoyant, 
allowing large masses of low entropy gas 
to be transported out beyond the cooling radius.
Successive cavities maintain a buoyant outflow that preserves 
the cluster gas temperature and gas fraction profiles and dramatically reduces 
the cooling rate onto the central black hole.
\end{abstract}

\date{\today}

\maketitle

\vspace{-0.4in}

\section{}

When a relatively small mass of gas accrets 
onto massive black holes in the cores of cluster and group-centered 
elliptical galaxies, enough energy can be released to solve the 
cooling flow problem: Why does the hot cluster gas radiate X-rays but 
does not appear to cool? 
Mass cooling rates must be reduced by at least an order of magnitude 
below those predicted by traditional cooling flows. 
While sufficient accretion energy is available, 
there is no general consensus regarding  
the mechanisms by which energy is delivered from the black hole  
to the hot gas. 
One of the major obstacles has been the observation that 
the gas temperature typically has a minimum at the center of the 
cluster atmosphere, 
just adjacent to the energy source at the black hole.

Fortunately, X-ray observations of cavities in the hot gas 
in (a minority of) clusters 
suggest that energy from the black holes 
can be transmitted from the central source by jets 
to both near and distant regions in the cluster gas. 
Radio emission from young X-ray cavities suggest that 
all cavities are inflated and supported largely by 
the pressure of relativistic cosmic rays 
that are created or deposited by jets. 
Cavities are not formed 
explosively but in a subsonic fashion,  
driving outward propagating shocks that 
increase the entropy lost by radiation in the local cluster gas. 
%Nevertheless, difficulties remain. 
It is usually assumed that the $PdV$ work 
done during cavity formation also compresses the ambient gas, 
helping to restore the thermal energy lost by radiation. 
However, while the entropy is increased by shocks around 
young cavities, most of the $PdV$ work is consumed in increasing the 
potential (not thermal) 
energy of cluster gas as gas moves out in the cluster potential 
to accommodate the cavity volume[5]. 
As cluster gas moves nearly adiabatically outward, its density decreases 
and it cools. 
For most, perhaps all, cavities 
this global expansive cooling offsets the 
heating by the cavity-produced shock. 
Furthermore, as shocks move out into the cluster gas 
the deposition of dissipative heating 
in the shock transition is not distributed in the cluster 
in proportion to the local radiative cooling.
Because the gas density profiles in clusters are generally 
flatter than $\rho \propto r^{-2}$, most wave energy is 
absorbed near the cluster centers. 
After a few Gyrs of heating by successive outward propagating 
waves, the gas temperature in the cluster core rises far above 
the temperatures observed and 
inflowing gas cooling from more distant 
regions of the cluster cools catastrophically just beyond the 
heated core[8]. 
Since cluster density profiles are not tuned to receive wave energy 
that balances local radiative losses, 
cavity (or other) shocks cannot be the dominant means of shutting down 
cooling flows.

These difficulties have led us to consider an alternative solution 
to the cooling flow problem in which low entropy gas near the 
cluster center is circulated outward by cosmic ray buoyancy to distant 
regions in the cluster without much disturbing the observed 
gas temperature profile.
%Successful explanations of the absence of cooling must be 
%simple and robust, particularly since 
%the cooling flow problem is common to 
%all astronomical scales -- galaxies, groups, and clusters.
For simplicity, we consider only the two most relevant 
cluster components, 
hot gas and (relativistic) cosmic rays. 
Since strong shocks are rare, as cavities evolve 
the combined pressure of thermal and cosmic ray gases must 
remain close to hydrostatic equilibrium in the cluster potential.
We assume that cosmic rays 
are deposited (or created) by jets at sites 
that become X-ray visible cavities. 
For simplicity we ignore 
non-adiabatic direct heating of the hot gas 
by cosmic rays via Columb heating or Alfven dissipation.

Cosmic rays and hot gas exchange momentum by means of small,
microgauss magnetic fields that are nearly frozen into the hot gas 
but which are too weak to influence the gas dynamics. 
Gradients in cosmic ray pressure cause X-ray cavities to form. 
In addition cosmic rays must be allowed to diffuse in the hot gas.
Diffusion is expected in part because of the notorious difficulty in 
confining a relativistic plasmas with magnetic fields 
and also because we expect synchrotron emission from the 
electron component of the cosmic rays to eventually 
evolve into radio lobes that can be much larger than the cavities. 
For simplicity we do not define 
in detail the particle nature of the cosmic rays -- electrons 
or protons -- and consider only their 
(relativistic) energy density. 
Our main interest is to understand how cosmic rays inflate 
X-ray cavities and buoyantly transport hot gas far from the 
cluster cores. 
In summary, cosmic rays are injected at cavity sites,
advected by the thermal gas,
diffuse relative to the gas and
have gradients that can act directly on the hot gas.

The rate that cosmic rays diffuse through the gas 
increases with cosmic ray particle energy, 
so the diffusion coefficient $\kappa$ we seek is a mean over the 
particle energy spectrum.
While $\kappa$ is difficult to derive, 
an approximate value can be determined from dimensional
considerations, $\kappa \propto$ length$^2$/time.
For example, spallation rates in the Milky Way indicate that 
cosmic rays reside in the disk plane 
having a scale height $z \approx 1$ kpc
for about 
$t_{spall} \approx 3 \times 10^7$ yrs, i.e. 
$\kappa \sim z^2/t_{spall} \sim 10^{28}$ cm$^2$/s, 
and this is very close to the values considered in detailed 
models of cosmic ray diffusion in the Milky Way.
Similarly,
cosmic rays must remain trapped in cavities for a typical 
cluster buoyancy time $t_{buoy} \sim 10^8$ yrs, 
but the appropriate scale length is not the radius of the cavity 
$r_{cav}$
but the thickness of the cavity wall, 
$\Delta r \sim 0.1r_{cav} \sim 1$ kpc, for which 
$\kappa \sim (\Delta r)^2/t_{buoy} \sim 10^{28}$ cm$^2$/s.
%While this is similar to the Milky Way value, 
It is also useful to explore larger rates, 
$\kappa \sim 10^{31}$ cm$^2$/s, 
to follow the progress of more energetic cosmic rays. 
For spherical cavities forming  
in a uniform thermal gas, the strength 
of circum-cavity shocks varies inversely with $\kappa$
and with the efficiency that 
cosmic rays diffuse through the cavity walls 
into the ambient gas[7]. 
Also $\kappa$ is likely to decrease in high density gas 
where the magnetic fields may be larger.

A major inspiration for the long term evolution 
of X-ray cavities has been the recent discovery of a 
narrow ($\sim 1$ kpc) $\sim 30$ kpc-long radial filament 
of cooler, slightly overdense thermal gas in the Virgo cluster[1].
Such radial features arise naturally from the ``splash'' 
following the formation of cavities in a gaseous atmosphere[2].  
Radial thermal filaments are driven by the confluence of gas 
flowing toward the bottom of buoyant cavities, 
which is relieved as gas squirts up through the cavity center 
and far beyond. 
(Cavitations following undersea explosions 
also produce intense, 
vertically directed flows above the sea surface.) 
The past history of Virgo is revealed in two very similar 
timescales: 
the dynamical time for the thermal filament to flow out from the 
cluster core, $\sim10^8$ yrs, and the 
synchrotron age $t_{sync}\sim10^8$ yrs  
of cosmic ray electrons in the large southern 
radio lobe which extends out to $r_{lobe} \sim$30-40 kpc[9].
Since the thermal filament lies right along the central 
diameter of the radio lobe, it is natural to conclude that 
they were formed by the same cavity event $\sim10^8$ yrs ago. 
The relevant diffusion coefficient for energetic cosmic rays 
that diffuse away from the cavity site,  
$\kappa \sim r_{lobe}^2/t_{sync} \sim 10^{29}$ cm$^2$/s, 
is reasonable. 
%and values $\sim 1000$ times lower to follow more strongly trapped 
%cosmic rays. 
Evolutionary models of a cavity inflated 
by diffusing cosmic rays at 
radius 10 kpc from the center of Virgo 
show that these two 
features could indeed have formed at the same time 
from a cavity that  
is no longer visible[6].
This is the first calculation that establishes a dynamical 
relationship between these large-scale cluster 
features observed at both X-ray and radio frequencies.

But the most important contribution of cosmic rays to 
cluster dynamics is not the short-term fireworks -- cavities, 
filaments, radio lobes, etc. -- but the 
long-lasting buoyant outflow of cluster gas to large radii. 
As cosmic rays diffuse into cluster gas, 
their pressure provides a small fraction of the total pressure 
required to maintain approximate local hydrostatic equilibrium. 
In such regions the gas pressure and density are slightly 
lowered and the cluster gas becomes naturally buoyant.
The local buoyant outflow velocity can easily exceed the 
(very small) cooling inflow velocity due to radiative losses.
By this means very large masses of low entropy gas can be slowly 
transported far from the galactic center before 
catastrophic radiative cooling occurs[5,4].

To illustrate this effect we explore how the 2D evolution 
of the Virgo cluster 
is affected by the presence of cosmic rays[4].
Consider first the response of the 
cluster without cosmic rays 
as it evolves by radiative losses 
away from its presently observed density and temperature
profiles, becoming a traditional cooling flow.
The left column in {\bf Figure 1} shows the initial 
configuration (solid lines) assumed to be in hydrostatic 
equilibrium and its density, temperature and 
pressure profiles after 3 Gyrs (long dashed lines). 
The dotted line shows the ratio of local gas entropy to 
its initial value $S/S_0$.
After 3 Gyrs the central temperature and entropy have dropped 
due to radiative losses and the gas density has 
adopted a central peak characteristic of traditional 
cooling flows. 
The large central cooling rate at this time, 
85 $M_{\odot}$ yr$^{-1}$, 
is incompatible with the absence of cooled gas in the cluster core.

However, the 2D flow is radically different in the presence of 
cosmic rays. 
Assume that successive cavities form 
every 200 Myrs 
at 10 kpc from the cluster center.
Each identical cavity is inflated in 20 Myrs 
by diffusing cosmic rays of total energy $8.2 \times 10^{58}$ ergs.
Two cosmic ray diffusion is given by
$\kappa = 10^{30} \min[1, n_{e0}/n_e]$ with 
$n_{e0} = 6 \times 10^{-6}$ (low $\kappa$) 
or $6 \times 10^{-3}$ (high $\kappa$).
The evolution of Virgo with cosmic rays 
after 3 Gyrs is shown in {\bf Figure 1} for each $\kappa(n_e)$.
Regardless of $\kappa$, it is seen that the gas density and 
temperature distributions are maintained close to the original profiles.
The entropy is enhanced locally near the cavity site (10 kpc), 
but $S/S_0$ is generally negative. 
Most of the low entropy gas near and within 10 kpc 
has been buoyantly transported out to 20-70 kpc 
(where $t_{cool}$ is much longer), 
but the cosmic ray pressure (short dashed lines) is 
generally small.

%%%%%%%%%%%%%%%%%%%%%%%%%%%%%%%%%%%%%%%%%%%%
%% Sample figure:
%%
%% The option [height=...] scales the picture to the given height,
%% without it it would be printed at its nominal size
%%%%%%%%%%%%%%%%%%%%%%%%%%%%%%%%%%%%%%%%%%%%                                       

\begin{figure}
  \includegraphics[height=.3\textheight]{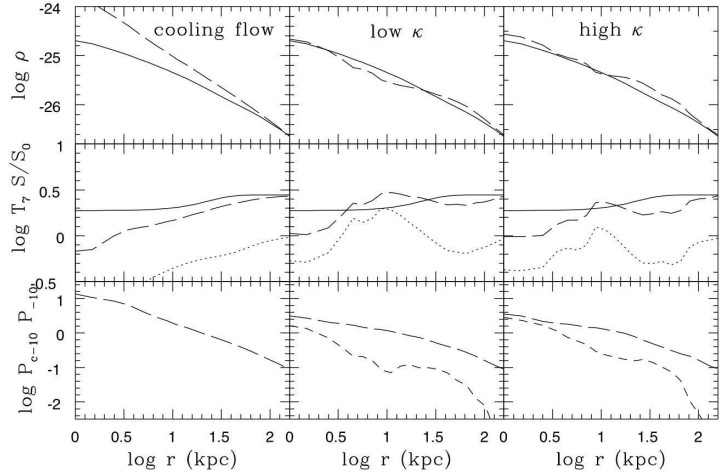}
  \caption{Evolution of Virgo after 3 Gyrs for pure cooling flow
  (left) and with cosmic ray feedback (center and right).
Dotted lines are $\log S/S_0$.}
\vspace{-0.2in}
\end{figure}

Most importantly, the central radiative cooling rates 
after 3 Gyrs have been reduced to 0.1 - 1 $M_{\odot}$ yr$^{-1}$, 
very far below the cooling flow value.
Low entropy gas that would have otherwise cooled has been made 
buoyant by cosmic rays that diffused through the cavity walls.
The time-averaged cosmic ray luminosity 
$3 \times 10^{43}$ erg s$^{-1}$ is modest, only equal to the 
observed X-ray luminosity within $\sim56$ kpc. Most of the 
energy that drives the buoyant outflow comes from the 
gravitational potential energy of the cluster.

This simple exploratory calculation suggests 
that the feedback deposition of cosmic rays into galaxy clusters 
provides a robust means of shutting down cooling flows. 
Moreover, the low central gas fraction observed in galaxy cluster cores 
can be created and maintained by cosmic ray feedback.
Finally, the transport of low-entropy gas to distant regions 
in the cluster is consistent with recent {\it Suzaku} observations
showing that the entropy in the outer regions of clusters is much 
lower than previously expected[3].
\centerline{}
% observe filament
\noindent
1. Forman, W., Jones, C., Churazov, E., et al. 2007, ApJ, 665, 1057\\
% cavity filament
\noindent
2. Gardini, A. 2004, A\&A 464, 143\\
% low entropy near rvir with Suzaku
\noindent
3. George, M. R., et al. 2009, MNRAS, 395, 657\\
% stopping cooling flows with cosmic rays
\noindent
4. Mathews, W. G., 2009, ApJ 695, L49\\
% energetics of x-ray cavities and radio lobes
\noindent
5. Mathews, W. G. \& Brighenti, F. 2008, ApJ 685, 128\\
% creation of filament and radio lobe in M87/Virgo:
\noindent
6. Mathews, W. G. \& Brighenti, F. 2008, ApJ 676, 880\\
% creation of x-ray cavities with cosmic rays
\noindent
7. Mathews, W. G. \& Brighenti, F. 2007, ApJ 660, 1137\\
% heating cooling flows with weak shock waves
8. Mathews, W. G., Faltenbacher, A., \& Brighenti, F. 2006, ApJ 638, 659\\
% radio lobe in M87/Virgo
\noindent
9. Owen, F. N., Eilek, J. A., \& Kassim, N. E. 2000, ApJ, 543, 611\\

\end{document}